%% file: 1bit-Asilomar_v1.tex
\documentclass[conference]{IEEEtran}
\IEEEoverridecommandlockouts
\usepackage{cite}
\usepackage{amsmath,amssymb,amsfonts}
\usepackage{upgreek}
\usepackage{algorithmic}
\usepackage{graphicx}
\usepackage{textcomp}
\usepackage{xcolor}
\usepackage{pgfplots}
\usepackage{multirow}
\usepackage{bbm}
\usepackage{vmr-symbols-vecbold}
\usepackage{siunitx}
\usepackage{microtype}
\usetikzlibrary{arrows,matrix,positioning,shapes.misc}

\DeclareMathOperator\rect{rect}

\definecolor{Red}{rgb}{1.0, 0, 0}

\renewcommand{\vec}[1]{\boldsymbol{\mathbf{#1}}}

\def\BibTeX{{\rm B\kern-.05em{\sc i\kern-.025em b}\kern-.08em
    T\kern-.1667em\lower.7ex\hbox{E}\kern-.125emX}}
\begin{document}

\title{Zero Crossing Modulation for Communication with Temporally Oversampled 1-Bit Quantization
\thanks{This work is supported by the Deutsche Forschungsgemeinschaft (DFG, German Research Foundation) in the Collaborative Research Center ``Highly Adaptive Energy-Efficient Computing'', SFB912, HAEC, Project-ID 164481002. Computations were performed at the ZIH at TU Dresden.}
} 

\author{\IEEEauthorblockN{Gerhard Fettweis\IEEEauthorrefmark{1}, Meik D\"orpinghaus\IEEEauthorrefmark{1}, Sandra Bender\IEEEauthorrefmark{1}, Lukas Landau\IEEEauthorrefmark{2}, Peter Neuhaus\IEEEauthorrefmark{1}, and Martin Schl\"uter\IEEEauthorrefmark{1}}
\IEEEauthorblockA{\IEEEauthorrefmark{1}Vodafone Chair Mobile Communications Systems, SFB 912 HAEC\\
Technische Universit\"at Dresden, 01062 Dresden, Germany\\
\{gerhard.fettweis, meik.doerpinghaus, sandra.bender, peter\_friedrich.neuhaus, martin.schlueter\}@tu-dresden.de}

\IEEEauthorblockA{\IEEEauthorrefmark{2}Center for Telecommunications Studies (CETUC), Pontifical Catholic University of Rio de Janeiro, RJ, Brazil\\
lukas.landau@cetuc.puc-rio.br}
}

\maketitle

\begin{abstract}
Today's communication systems typically use high resolution analog-to-digital converters (ADCs). However, considering future communication systems with data rates in the order of \SI{100}{\giga bit/s} the ADC power consumption becomes a major factor due to the high sampling rates. A promising alternative are receivers based on 1-bit quantization and oversampling w.r.t.\ the signal bandwidth. Such an approach requires a redesign of modulation, receiver synchronization, and demapping. A zero crossing modulation is a natural choice as the information needs to be carried in the zero crossing time instants. The present paper provides an overview on zero crossing modulation, achievable rates, sequence mapping and demapping, 1-bit based channel parameter estimation, and continuous phase modulation as an alternative zero crossing modulation scheme. 
\end{abstract}

\section{Introduction}
For very high data rate communications with data rates in the order of $100~\mathrm{Gbit/s}$ the power consumption of the ADC becomes a major factor. The reason is that the large bandwidth of tenth of $\mathrm{GHz}$ required for such high data rates implies very high sampling frequencies, which limit the achievable quantization resolution given a fixed power budget of the ADC. In this regard, the key point is that for sampling frequencies beyond $100~\mathrm{MHz}$ the energy per conversion step begins to increase linearly with the frequency, while for smaller sampling rates the energy per conversion step is independent of the sampling rate \cite{murmann2013energy,Murmann_ADC}. One promising alternative is the use of 1-bit quantization with temporal oversampling w.r.t.\ the Nyquist rate of the received signal, i.e., spending resolution in the time domain instead of the amplitude domain. In this regard, 1-bit quantization is advantageous as it is simple to realize since it requires no automatic gain control and no highly linear analog signal processing. 1-bit quantization is robust against amplitude uncertainties such that not much headroom for sophisticated processing in the amplitude domain is needed. Thus, besides the simpler circuitry without an automatic gain control, 1-bit quantization allows to use energy efficient circuits with supply voltages smaller than $1~\mathrm{V}$ \cite{staszewski2012digitally}. Moreover, time-domain resolution is becoming superior to amplitude-domain resolution in modern nanometer-scale CMOS processes \cite{staszewski2012digitally,Ziabakhsh18}. For this reason, we study wireless communications with temporally oversampled 1-bit quantization.\looseness-1

As different signal amplitudes cannot be distinguished after 1-bit quantization, the information needs to be carried in the zero crossing time instants of the transmit signal and ADC resolution is spend in time rather than in amplitude. Optimal communication over such a channel requires an adapted modulation and signaling scheme. A natural choice to communicate over the resulting channel is a so called \emph{zero crossing modulation}. Fig.~\ref{Zero-Crossing-Signal} shows a zero crossing modulated transmit signal where the sequence of distances between the zero crossings encodes the information. Such transmit signals can be generated by combining faster-than-Nyquist (FTN) signaling \cite{mazo1975faster} with runlength coding \cite{immink1990runlength}. While FTN signaling introduces intersymbol interference (ISI), this interference can be well controlled by an appropriate choice of the minimum runlength constraint of the runlength code. The FTN signaling rate and correspondingly the oversampling rate w.r.t.\ the Nyquist rate at the receiver determine the resolution on which the zero crossing time instants can be chosen and, thus, have a strong impact on the achievable rate.

In this paper, we give an overview on communication with temporally oversampled 1-bit quantization. We discuss the spectral efficiency with different modulation schemes and different oversampling rates, which motivates the use of runlength coding in combination with FTN signaling. Subsequently, we also discuss some aspects on the actual design of such a communication system. We describe the mapping of the code bits onto the runlength limited (RLL) sequences and equalization and demapping at the receiver. Another key challenge is channel estimation and synchronization. As the 1-bit quantization is a highly nonlinear operation standard algorithms for channel parameter estimation and synchronization cannot be used. As a first step to the design of actual channel parameter estimation algorithms, we studied the impact of the 1-bit quantization on the Cram\'er-Rao lower bound (CRLB) for the estimation of phase and frequency offsets, giving some important insight into the actual receiver design. Moreover, we use the CRLB to benchmark a least squares phase estimator. Finally, we also discuss the use of continuous phase modulation (CPM) for oversampled 1-bit quantization based receivers, for which sequences of zero-crossings can be designed that allow recovery of higher order CPM modulation schemes despite 1-bit quantization. Additionally, as a constant envelope scheme, this approach relaxes the linearity requirements on the transmitter side and, thus, enables the use of more energy efficient power amplifiers.

\emph{Notation:} Random quantities are denoted by sans serif letters, e.g., $\mathsf{a}$, while $a$ denotes a deterministic quantity.

\section{System Model}\label{SectSysModel}
As motivated before we propose to use a zero crossing modulation (ZXM) which is naturally matched to a receiver using 1-bit quantization with temporal oversampling. Such a receiver can only resolve the sign of the received signal at the sampling time instances, i.e., effectively it can retrieve the temporal distances between the zero crossings of the received signal. Thus, the ZXM needs to convey the information in the distances between the zero crossings. 

A ZXM transmit signal can be created utilizing RLL sequences \cite{immink1990runlength}, which can be generated from $(d,k)$-sequences. In a $(d,k)$-sequence a $1$ is followed by at least $d$ and at most $k$ 0s. The limitation of the runlength by the $k$ constraint can be important for synchronization at the receiver. However, in the present work we assume $k=\infty$ which, for a chosen value of $d$, maximizes the entropy rate of the generated RLL sequences. The generation of the $(d,k)$-sequences can be represented based on a finite-state machine as shown in Fig.~\ref{fig:rll_fsm}. The $(d,k)$-sequences are transformed into RLL sequences by non return to zero inversion (NRZI) encoding as shown by the following example with $d=1$:
\begin{IEEEeqnarray}{rCL}
\begin{array}{lcccccccccc}
(d,k)\textrm{-seq.} & [\hdots  & 1 & 0 & 0 & 0 & 1 & 0 & 1 & 0 & \hdots]\\
\textrm{RLL-seq.} & [\hdots  & 1 & 1 & 1 & 1 & -1 & -1 & 1 & 1 & \hdots]
\end{array}\nonumber
\end{IEEEeqnarray}
Hence, the minimum distance between two zero crossings can be controlled by the $d$ constraint. 

Two such runlength limited sequences are used to generate a complex valued transmit symbol sequence, which is given by\looseness-1
\begin{IEEEeqnarray}{rCL}
\bm{\mathsf{x}}^N&=&\frac{1}{\sqrt{2}}\left(\bm{\mathsf{a}}^N+j\bm{\mathsf{b}}^N\right)
\end{IEEEeqnarray}
where the elements of the vectors $\bm{\mathsf{a}}^N$ and $\bm{\mathsf{b}}^N$ are given by two independent real valued RLL sequences of length $N$. Thus, the elements $\mathsf{x}_n$ of $\bm{\mathsf{x}}^N$ are QPSK symbols with $\mathsf{x}_n\in \mathcal{X} = \left\{\frac{1+j}{\sqrt{2}},\frac{1-j}{\sqrt{2}},\frac{-1+j}{\sqrt{2}},\frac{1+j}{\sqrt{2}}\right\}$.

Using transmit symbol sequences according to $\bm{\mathsf{x}}^N$ in combination with FTN signaling, we generate the continuous-time channel input signal
\begin{IEEEeqnarray}{rCL}
\mathsf{x}(t)&=&\sum_{n=-\frac{N-1}{2}}^{\frac{N-1}{2}}\mathsf{x}_n h\left(t-\frac{nT}{M_{\mathrm{Tx}}}\right)\label{GeneralInputSignal}
\end{IEEEeqnarray}
with the symbol rate $\frac{M_{\mathrm{Tx}}}{T}$ where $T$ is the unit time interval and $M_{\mathrm{Tx}}$ is the FTN signaling factor.\footnote{In case $h(t)$ is a Nyquist pulse, $T$ is the length of the Nyquist interval.} Finally, $h(t)$ is the impulse response of the transmit filter. Unless otherwise mentioned we use a cosine pulse as transmit filter given by
\begin{IEEEeqnarray}{rCL}
h_{\textrm{cos}}(t)&=&\left\{\begin{array}{ll}
\sqrt{\frac{1}{3T}}\left(1-\cos\left(2\pi\frac{t}{2T}\right)\right) & 0\le t<2T\\
0 & \textrm{otherwise}
\end{array}\right. .\label{TXFilterCos}
\end{IEEEeqnarray}
Using such a transmit filter with a relatively short impulse response has the advantage of enabling trellis based receivers with a limited complexity.

Using FTN signaling with $M_{\mathrm{Tx}}>1$ allows us to increase the resolution of the time grid on which we can place the zero crossings. This counteracts the decrease of the achievable rate caused by the $d$ constraint in the RLL sequence and allows to increase the achievable rate. Note that FTN signaling introduces ISI, which however can be well controlled by an appropriate choice of the minimum runlength $d$. 

\begin{figure}
\centerline{\includegraphics[width=\columnwidth]{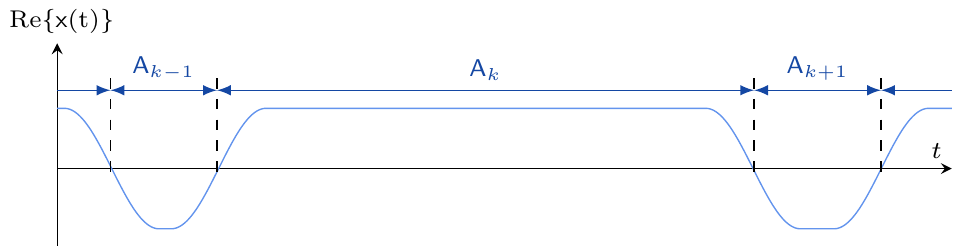}}
\caption{Schematic example of a zero crossing modulated transmit signal generated by runlength coding and FTN.}
\label{Zero-Crossing-Signal}
\end{figure}

A schematic view of a zero crossing modulated transmit signal is given in Fig.~\ref{Zero-Crossing-Signal}. Here $\mathsf{A}_k=\frac{T}{M_{\textrm{Tx}}}\mathsf{l}_k$ with the elements of the sequence $\{\mathsf{l}_k\}$ being the runlengths in the RLL sequence.

We consider an additive white Gaussian noise channel, such that the signal at the output of the receive filter is given by
\begin{IEEEeqnarray}{rCL}
\mathsf{z}(t)&=&\int_{-\infty}^{\infty}\left(\mathsf{x}(\tau)+\mathsf{n}(\tau)\right)h_{\textrm{Rx}}(t-\tau)\mathrm{d}\tau\label{RxSignalContTime}
\end{IEEEeqnarray}
with the receive filter $h_{\textrm{Rx}}(t)$. Unless otherwise stated, we use an integrator over the time interval $T/M_{\textrm{Tx}}$ as receive filter, which corresponds to the integrator in an integrate-and-dump receiver. Its impulse response is given by
\begin{IEEEeqnarray}{rCL}
h_{\textrm{Rx}}(t)&=&\left\{\begin{array}{ll}
\sqrt{\frac{M_{\textrm{Tx}}}{T}} & 0\le t<\frac{T}{M_{\textrm{Tx}}}\\
0 & \textrm{otherwise}\end{array}\right. .\label{h_rx}
\end{IEEEeqnarray}
Also here the short impulse response is helpful to limit the complexity of trellis based sequence detection. Moreover, $\mathsf{n}(t)$ is a circularly symmetric complex white Gaussian noise process with power spectral density $N_0$. Note that we assume here perfect timing, phase, and frequency synchronization. Later in Section~\ref{Sect_ChannelParameterEstimation}, we will discuss the channel parameter estimation.\looseness-1

The signal $\mathsf{z}(t)$ is sampled at rate $1/T_s$ yielding the following signal before 1-bit quantization
\begin{IEEEeqnarray}{rCL}
\mathsf{r}_k&=&\sum_{n=-\frac{N-1}{2}}^{\frac{N-1}{2}}\mathsf{x}_ng\left(kT_s-\frac{nT}{M_{\textrm{Tx}}}\right)+\mathsf{n}_k\ \label{RxSignalbeforeADC}\ 
\end{IEEEeqnarray}
with 
\begin{IEEEeqnarray}{rCL}
g(t)&=&\int_{-\infty}^{\infty}h(\tau)h_{\textrm{Rx}}(t-\tau)\mathrm{d}\tau\\
\mathsf{n}_k&=&\int_{-\infty}^{\infty}\mathsf{n}(\tau)h_{\textrm{Rx}}(kT_s-\tau)\mathrm{d}\tau.
\end{IEEEeqnarray}
Moreover, we define the oversampling factor as $M=\frac{T}{T_s M_{\textrm{Tx}}}$. Note that with the receive filter in (\ref{h_rx}) for $M>1$ the noise samples $\mathsf{n}_k$ are correlated.

Finally, the signal at the output of the 1-bit ADC is given by\looseness-1 
\begin{IEEEeqnarray}{rCL}
\mathsf{y}_k&=&\mathrm{csign(\mathsf{r}_k)}=\mathrm{sign}(\mathrm{Re}\{\mathsf{r}_k\})+j\cdot \mathrm{sign}(\mathrm{Im}\{\mathsf{r}_k\}) \label{eq:Qunatizer1bit}
\end{IEEEeqnarray}
with the signum function $\mathrm{sign}(x)=1$ if $x>0$ and $\mathrm{sign}(x)=-1$ if $x\le 0$.

As the transmit signal is not strictly bandlimited, we use the following SNR definition 
\begin{IEEEeqnarray}{rCL}
\textrm{SNR}&=&\frac{\lim_{T'\rightarrow\infty}\frac{1}{T'}\int_{T'}|\mathsf{x}(t)|^2\mathrm{d}t}{N_0 B_{\textrm{90\%}}}
\end{IEEEeqnarray}
where $B_{\textrm{90\%}}$ denotes the $90\%$ power containment bandwidth of the transmit signal $\mathsf{x}(t)$.

\section{Spectral Efficiency}
To understand the limits on the achievable data rate for such a communication system with 1-bit quantization and temporal oversampling we have studied its spectral efficiency. In the following, we will mainly give an overview on the spectral efficiency when using RLL sequences and FTN signaling as described in Section~\ref{SectSysModel}. For comparison we will also show the spectral efficiency with FTN signaling for i) transmit symbol sequences $\{\mathsf{x}_n\}$ consisting of independently uniformly distributed (i.u.d.) transmit symbols and ii) for transmit symbol sequences where the elements $\mathsf{x}_n$ are the output of a Markovian source in which the transition probabilities are optimized by the Blahut-Arimoto algorithm such that the achievable rate is maximized. For details on this approach the reader is referred to \cite{landauEurasip2018}. The results presented in this section have been published in \cite{landau2015communications} and \cite{landauEurasip2018}, where more details can be found.

\subsection{Lower-Bounding the Achievable Rate}\label{subsec:AchRate}
The achievable rate can be evaluated based on the following auxiliary channel lower bound \cite{arnold2006simulation}
\begin{IEEEeqnarray}{rCL}
I'(\bm{\mathsf{x}};\bm{\mathsf{y}})&=&\lim_{N\rightarrow\infty}\frac{1}{N}I\left(\bm{\mathsf{x}}^{N};\bm{\mathsf{y}}^{N}\right)\nonumber\\
&\ge&\frac{1}{N}\left(-\log_2 W\left(\mathbf{y}^{N}\right)+\log_2 W\left(\mathbf{y}^{N}|\mathbf{x}^{N}\right)\right)\!.\label{AuxLB}
\end{IEEEeqnarray}
Here, $\bm{\mathsf{y}}^{N}=[\bm{\mathsf{y}}_1^T,\hdots,\bm{\mathsf{y}}_N^T]^T$ is an $M N$-dimensional vector with $\bm{\mathsf{y}}_n$ being vectors of dimension $M$ containing the receive samples $\mathsf{y}_k$ at the output of the 1-bit quantizer belonging to the transmit symbol $\mathsf{x}_n$. Note that in case the receiver samples at the FTN signaling rate, i.e., $M=1$, the length of $\bm{\mathsf{x}}^{N}$ and  $\bm{\mathsf{y}}^{N}$ are both $N$.

Moreover, $W(\cdot)$ is an auxiliary channel based on the approximation $P(\mathbf{y}_k|\mathbf{y}^{k-1},\mathbf{x}^{k})\approx P(\mathbf{y}_k|\mathbf{x}_{k-L}^{k})$ with $\mathbf{x}_{k-L}^{k}=[x_{k-L},\hdots,x_{k}]^T$. Here, $L+1$ is the length of the impulse response of the effective channel containing the transmit and the receive filter in terms of $T/M_{\textrm{Tx}}$. To lower-bound the achievable rate, the RHS of (\ref{AuxLB}) is numerically evaluated for very large $N$ based on realizations of $\bm{\mathsf{x}}^{N}$ and $\bm{\mathsf{y}}^{N}$ denoted by $\mathbf{x}^{N}$ and $\mathbf{y}^{N}$.

The spectral efficiency is given by
\begin{IEEEeqnarray}{rCL}
\textrm{SE}&=&\frac{I'(\bm{\mathsf{x}};\bm{\mathsf{y}})M_{\textrm{Tx}}}{T B_{\textrm{90\%}}}
\end{IEEEeqnarray}
as $I'(\bm{\mathsf{x}};\bm{\mathsf{y}})$ is the achievable rate per time $\frac{T}{M_{\textrm{Tx}}}$. For more details on the evaluation of the lower bound in (\ref{AuxLB}) the reader is referred to \cite{landauEurasip2018,landau2015communications}.

\subsection{Runlength Limited Sequences}\label{Sect_SpecEff_RLL}
\begin{figure}
	\resizebox{\columnwidth}{!}{
		\centering
		\begin{tikzpicture}
		\tikzset{
			myCirc/.style = {draw, circle, minimum width=1.3cm}
		}
		\node[myCirc] (n1-1) at (0,0) {$0$};
		\node[myCirc] (n2-1) at (2.5,0) {$1$};
		\node[] (nHelp-1) at (4.5,0) {$\dots$};
		\node[myCirc] (n3-1) at (6.5,0) {$(\mathrm{d}-1)$};
		\node[myCirc] (n4-1) at (9,0) {$\mathrm{d}$};
		
		\draw[->,above] (n1-1) -- node {$0$} (n2-1);
		\draw[->,above] (n2-1) -- node {$0$} (nHelp-1);
		\draw[->,above] (nHelp-1) -- node {$0$} (n3-1);
		\draw[->,above] (n3-1) -- node {$0$} (n4-1);
		\path (n4-1) edge [->,out=-110,in=-70,looseness=0.2] node[above] {$1$} (n1-1);
		\path (n4-1) edge [->,out=-30,in=30,looseness=5] node[right] {$0$} (n4-1);
		
		\end{tikzpicture}}
	\caption{Definition of a $(d,k)$-sequence with $k=\infty$ as a finite-state machine.}
	\label{fig:rll_fsm}
\end{figure}
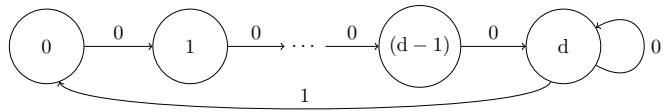

As we are interested in the maximal spectral efficiency, we use maximum entropy runlength sequences for $\bm{\mathsf{a}}^N$ and $\bm{\mathsf{b}}^N$, i.e., for the real and imaginary component of $\bm{\mathsf{x}}^N$. For a given $d$ and $k=\infty$ the maximum entropy rate of a runlength sequence is given by \cite{immink1990runlength,shannon1948mathematical}
\begin{IEEEeqnarray}{rCL}
\label{eq:def_Hmax}
H_{\textrm{max}}&=&\lim_{N\rightarrow\infty}\frac{1}{N}\log_2\sum_{i,j}\left[\mathbf{D}^N\right]_{i,j}=\log_2 (\lambda)
\end{IEEEeqnarray}
where $\mathbf{D}$ is the adjacency matrix of the finite-state machine in Fig.~\ref{fig:rll_fsm} and $\lambda$ is its largest eigenvalue. Exemplarily, the adjacency matrices for $d=1$ and $d=2$ are given by
\begin{IEEEeqnarray}{rCL}
\mathbf{D}_{d=1}&=&\left(\begin{array}{cc}
0 & 1 \\
1 &1 
\end{array}\right), \quad 
\mathbf{D}_{d=2}=\left(\begin{array}{ccc}
0 & 1 & 0\\
0 & 0 & 1\\
1 & 0 & 1
\end{array}\right)
\end{IEEEeqnarray}
where the rows correspond to the current state and the columns to the following state. The transition probabilities for the maximum entropy runlength sequences are given by \cite[eq. (2.23)]{immink2004codes}\looseness-1
\begin{IEEEeqnarray}{rCL}
[\mathbf{P}]_{i,j}&=&\frac{u_j}{u_i}\cdot\frac{[\mathbf{D}]_{i,j}}{\lambda}
\end{IEEEeqnarray}
with $u_i$ and $u_j$ being the $i$-th and the $j$-th entry of the right-hand eigenvector belonging to the eigenvalue $\lambda$. As with increasing $d$ the randomness in the runlength sequence decreases, the maximum entropy rate decreases with $d$, see Table~\ref{MxEntropRLL}. However, the overall data rate is increased due to FTN signaling, i.e., $M_{\textrm{Tx}}>1$, beyond \SI{2}{bit/s/Hz}.

\begin{table}
\center
\caption{Maximum entropy rate of $d$-constrained runlength sequences}\begin{tabular}{|l|c|c|c|}
\hline
minimum runlength constraint $d$ & 1 & 2 & 3 \\
\hline 
$H_{\textrm{max}}$ [bit/symbol] & 0.6942 & 0.5515 & 0.4650 \\
\hline
\end{tabular}
\label{MxEntropRLL}
\end{table}

\subsection{Numerical Results}
\begin{figure}
\center	\includegraphics[width=0.48\textwidth]{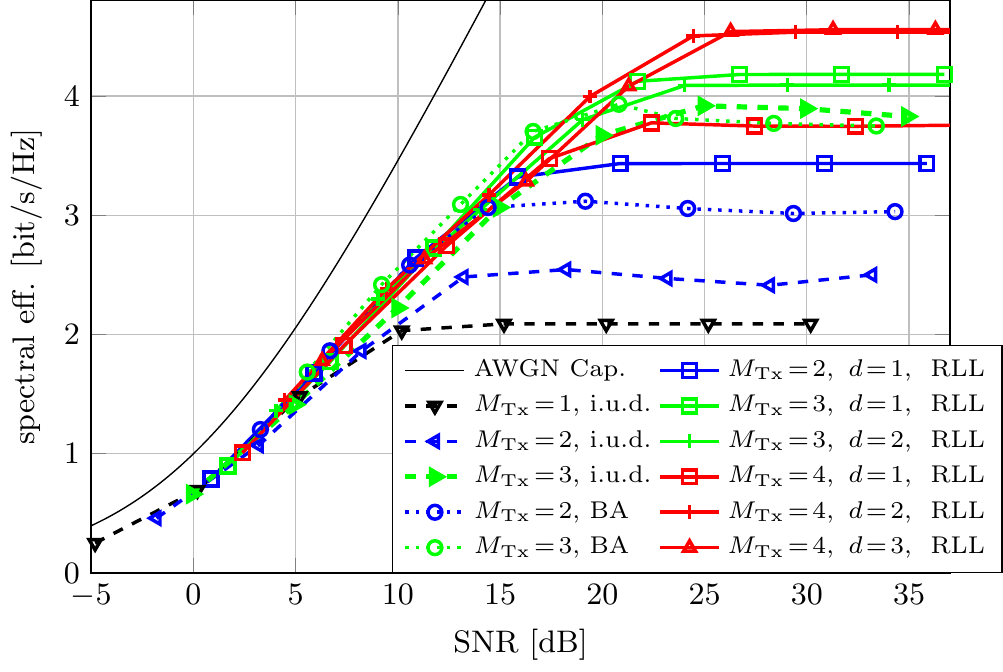}\vspace{-0.1cm}
	\caption{Spectral efficiency versus SNR for different sequence designs with FTN signaling rate $M_{\textrm{Tx}}$; $M=1$; `i.u.d.' independent uniformly distributed input symbols, `RLL' runlength coding with minimum runlength $d$, and `BA' Markov source sequences with optimized transition probabilities using the Blahut-Arimoto algorithm. The figure is an extended version of \cite{landau2015communications}.}
	\label{fig:AchievableRateOversamp}
\end{figure}

Fig.~\ref{fig:AchievableRateOversamp} shows the spectral efficiency based on the $90\%$ power containment bandwidth for the different transmit symbol sequence designs. Note that for the here considered case of $M=1$ and the receive filter in (\ref{h_rx}) the lower bound in (\ref{AuxLB}) is tight as $P(\mathbf{y}_k|\mathbf{y}^{k-1},\mathbf{x}^{k})= P(\mathbf{y}_k|\mathbf{x}_{k-L}^{k})$. 

Regarding the spectral efficiency with RLL sequences and FTN signaling the following can be observed. While without FTN signaling ($M_{\textrm{Tx}}=1$)  and with i.u.d.\ sequences the spectral efficiency saturates at round about \SI{2}{bit/s/Hz} as expected\footnote{With $M_{\textrm{Tx}}=1$ the spectral efficiency is slightly higher than \SI{2}{bit/s/Hz} as we consider the $90\%$ power containment bandwidth because the transmit signal is not strictly bandlimited.}, FTN signaling enables significantly higher spectral efficiencies. For the shown FTN signaling rates the spectral efficiency increases with $M_{\textrm{Tx}}$. However, the runlength constraint $d$ needs to be chosen appropriately. In general $d$ needs to be increased with $M_{\textrm{Tx}}$ to limit the ISI. I.e., for $M_{\textrm{Tx}}=4$ choosing $d=1$ leads to a significantly lower spectral efficiency than $d=2$. However, $d$ should also not be chosen too large, as this limits the entropy of the RLL sequence too strongly as can, e.g., be observed for the case  $M_{\textrm{Tx}}=3$ and $d=2$. On the other hand, the bandwidth also decreases with increasing $d$.

Moreover, Fig.~\ref{fig:AchievableRateOversamp} shows that FTN signaling using runlength coding yields a higher spectral efficiency than using i.u.d.\ transmit symbol sequences. For comparison we also considered FTN signaling with transmit symbol sequences generated by a Markov source with transition probabilities optimized by the Blahut-Arimoto algorithm, denoted by `BA'. Compared to this approach, runlength coding yields a spectral efficiency that is mostly superior or at least similar. This is due to the fact that the optimization is performed w.r.t.\ the achievable rate in (\ref{AuxLB}) and not w.r.t.\ the spectral efficiency, which is not equivalent due to the influence of the optimization on the $90\%$ power containment bandwidth. 

In \cite{LandauCommLett17} we have also studied the spectral efficiency in case of a strictly bandlimited channel, showing that also in this case temporal oversampling w.r.t.\ the Nyquist rate enables significant gains in the achievable rate. This supports the statement that gains presented in Fig.~\ref{fig:AchievableRateOversamp} while using the $90\%$ power containment bandwidth are not mainly due to the out of band transmit signal energy.

Overall, this study shows that runlength coding in combination with FTN signaling can significantly increase the spectral efficiency and is an appropriate sequence design approach. Note that the runlength coding on the first hand leads to a decrease of the achievable rate. However, in combination with FTN signaling it allows to reduce the ISI significantly and, thus, overall results in an increase of the spectral efficiency. These numerical results on the spectral efficiency are supported by analytical bounds on the spectral efficiency for a continuous-time channel in \cite{bender2017achievable} which corresponds to the case of  $M_{\textrm{Tx}}\rightarrow\infty$ and $T_s\rightarrow 0$.

\section{Sequence Mapping and Demapping}
\begin{figure}
	\tikzset{
		block/.style = {draw, fill=white, rectangle,
			minimum height=3em, minimum width=2cm},
		circ/.style = {draw, fill=white, circle, inner sep=0pt},
		input/.style = {coordinate},
		output/.style = {coordinate},
		pinstyle/.style = {pin edge={to-,t,black}}
	}
	\centering
	\resizebox{\columnwidth}{!}{
		\begin{tikzpicture}[auto, draw=black, node distance=2cm, >=latex']
		\node (b) at (-9.75,0) {\Large $b_l$};
		\node[block] (fecEnc) at (-8,0) {\Large FEC enc.};
		\node[block] (rllEnc) at (-5,0) {\Large RLL enc.};
		\node[block] (nrziEnc) at (-2.0,0) {\Large NRZI enc.};
		\node[block] (dac) at (1,0) {\Large DAC};
		\node (dac_label) at (1,+1) {\Large signaling rate $\frac{M_\mathrm{Tx}}{T}$};
		\node (x_n) at (2.65,0.3) {\Large $\mathsf{x}_n$};
		\node[block] (f) at (4.3,0) {\Large $h(t)$};
		\node (x_cont) at (6.3,0.35) {\Large $\mathsf{x}(t)$};
		\node (n0) at (6.1,-1.25) {\Large $\mathsf{n}(t)$};
		\node[circ] (a0) at (7.1,-1.25) {$+$};
		\node[block] (g0) at (4.3,-2.5) {\Large $h_\mathrm{Rx}(t)$};
		\node (z_cont) at (2.75,-2.15) {\Large $\mathsf{z}(t)$};
		\node[block] (adc0) at (1,-2.5) {\Large 1-bit ADC};
		\node (adc_label0) at (1,-1.5) {\Large sampling rate $\frac{1}{T_\mathrm{s}}$};
		\node (y_k) at (-0.525,-2.15) {\Large $\mathsf{y}_k$};
		\node[block] (bcjr10) at (-2.0,-2.5) {\Large Equalizer};
		\node (xhat_n) at (-3.45,-2.15) {\Large $\hat{\mathsf{x}}_n$};
		\node[block] (bcjr20) at (-5,-2.5) {\Large RLL dec.};
		\node[block] (fecDec0) at (-8,-2.5) {\Large FEC dec.};
		\node (r0) at (-9.75,-2.5) {\Large $\hat{b}_{l}$};
		
		\draw[->] (b) -- node {} (fecEnc);	
		\draw[->] (fecEnc) -- node {} (rllEnc);
		\draw[->] (rllEnc) -- node {} (nrziEnc);
		\draw[->] (nrziEnc) -- node {} (dac);
		\draw[->] (dac) -- node {} (f);	
		\draw[->] (f) -| (a0);
		\draw[->] (n0) -- (a0);
		\draw[->] (a0) |- (g0);
		\draw[->] (g0) -- (adc0);
		\draw[->] (adc0) -- (bcjr10);
		\draw[->] (bcjr10) -- (bcjr20);
		\draw[->] (bcjr20) -- (fecDec0);
		\draw[->] (fecDec0) -- (r0);
		
		\draw [color=black,dashed](-10.25,1.5) rectangle (5.55,-0.75);
		\node at (-10.25,1.25) [below=0mm, right=0mm] {\Large Transmitter};

		\draw [color=black,dashed](-10.25,-1) rectangle (5.55,-3.25);
		\node at (-10.25,-1.25) [below=0mm, right=0mm] {\Large Receiver};
		
		\end{tikzpicture}}
	\caption{System model.}
	\label{fig:system_model}
\end{figure}
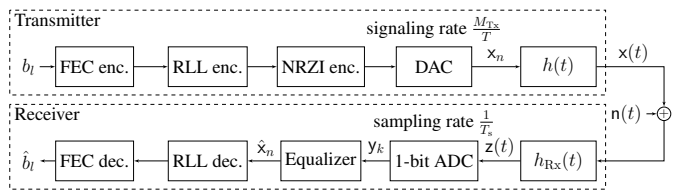

We consider the system model in Fig.~\ref{fig:system_model}.
In order to generate the transmit signal, it is necessary to map the encoded information bits onto an RLL sequence.
As described in Section~\ref{SectSysModel}, we propose to employ independently modulated RLL sequences for the real and imaginary part of the transmit signal.
Block codes present a widely used low-complexity encoding method.
A survey of different encoding techniques can be found in \cite{immink1990runlength}.
The maximum rate $R_\mathrm{RLL}$ of any practical encoding scheme is bounded by $R_\mathrm{RLL} \leq H_{\textrm{max}}$, where $H_{\textrm{max}}$ is given in \eqref{eq:def_Hmax}, and its efficiency can be measured by $\eta = \frac{R_\mathrm{RLL}}{H_{\textrm{max}}}$.

In this work we utilize a block code from \cite[Tab.~4]{immink1990runlength} which maps $3$ information bits onto a $(d=1,k=\infty)$ sequence of length $5$.
Hence, the code has a rate of $R_\mathrm{RLL}=0.6$ and an efficiency of $\eta \approx 0.86$ (cf. Tab.~\ref{MxEntropRLL}).

\subsection{Equalizer}
The system model in \eqref{eq:Qunatizer1bit} corresponds to a 1-bit quantized ISI channel with correlated noise for $M>1$.
The equalizer aims to remove the ISI and to provide reliability information, i.e., soft information, for each transmitted RLL symbol by performing maximum a posteriori (MAP) symbol detection, i.e.,\looseness-1
\begin{equation}
    \hat{x}_n = \arg \max_{x_n \in \mathcal{X}} P( x_n \vert \mathbf{y}^N), \quad n \in \lbrace 1, \ldots, N \rbrace.
\end{equation}
MAP symbol detection can be implemented efficiently by the BCJR algorithm \cite{bcjr1974} if a suitable factorization of the a posteriori probability $P( \mathbf{x}^N \vert \mathbf{y}^N )$ is found. However, such a factorization is unknown in case of correlated noise, i.e., $M>1$, and in that case the noise correlation has to be ignored in order to obtain a factorization.
The BCJR algorithm works on a trellis whose size, i.e., the number of states, is determined by the ISI memory length introduced by the transmit and receive filter.
The number of trellis states and the number of transitions can however be reduced by omitting states and transitions which violate the RLL constraint.
Note that the equalizer yields soft information despite 1-bit quantization.
The reader is referred to \cite{Neuhaus2019} for more details on an equalizer implementation.\looseness-1

\subsection{Soft-Input Soft-Output RLL Demapping}
After equalization the receiver performs soft-input soft-output demapping/decoding of the RLL sequence.
We propose to implement joint NRZI and RLL demapping, by extending the RLL block code to also include the NRZI mapping.
Then soft-input soft-output demapping can be implemented by first computing the probability of each code word in the block code and subsequently computing the probability of each input bit based on the code word probabilities.
Note that this procedure is similar to soft demapping of, e.g., QAM constellations.

\subsection{Numerical Results}
The performance of the proposed system is evaluated for an AWGN channel.
As forward error correction (FEC) scheme we utilize base graph 1 of the 5G NR LDPC code, including CRC coding, code block segmentation, rate matching and rate recovery (which include interleaving and deinterleaving).
The code block size of the channel code is chosen to the maximum of $N_\mathrm{FEC}=25344$.

\begin{figure}
	\centering
	\input{fig_eval_bler.tex}\vspace{-0.1cm}
	\caption{Coded performance evaluation for $M_\mathrm{Tx}=2$ using 5G NR LDPC codes. The corresponding spectral efficiencies are approximately \SI{2.23}{bit/s/Hz}, \SI{2.37}{bit/s/Hz}, and  \SI{2.64}{bit/s/Hz} for the code rates $R_\mathrm{FEC}$ of $3/4$, $4/5$, and $8/9$, respectively, where the spectral efficiencies are computed using the 90\% power containment bandwidth of the max-entropic RLL sequences discussed in Sect.~\ref{Sect_SpecEff_RLL}.} 
	\label{fig:eval_bler}
\end{figure}
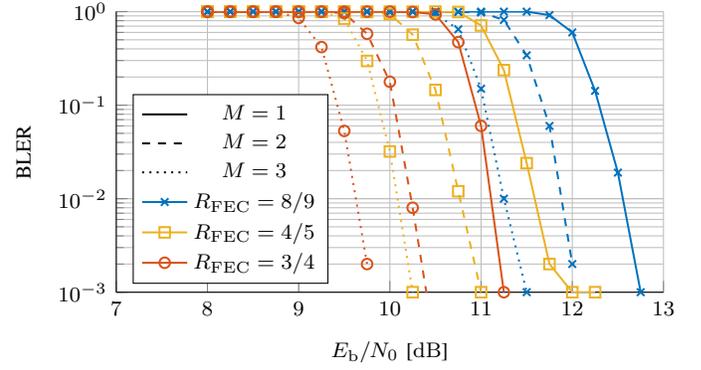

A coded performance evaluation w.r.t.\ block-error rate (BLER) is depicted in Fig.~\ref{fig:eval_bler}.
The system behaves similar for all considered code rates.
It can be observed that oversampling increases the performance significantly.
Each integer increase in $M$ result in approx. \SIrange{0.5}{0.8}{\decibel} $E_\mathrm{b}/N_0$ gain at a BLER of $10^{-2}$.

\section{Channel Parameter Estimation}\label{Sect_ChannelParameterEstimation}
The channel parameter estimation and synchronization based on 1-bit quantized received signals poses new challenges as standard estimation and synchronization algorithms cannot be applied. In this regard, consider that in the unsynchronized case the received signal after receive filtering and sampling at rate $1/T_s$ is given by 
\begin{IEEEeqnarray}{rCL}
\mathsf{r}_k&=&\sum_{n=-\frac{N-1}{2}}^{\frac{N-1}{2}}\mathsf{x}_ng\left(kT_s-\frac{nT}{M_{\textrm{Tx}}}-\epsilon T\right)e^{j(\Omega kT_s+\phi+\varphi_k)}+\mathsf{n}_k\nonumber\\
&=& \mathsf{u}_k e^{j(\Omega kT_s+\phi+\varphi_k)} + \mathsf{n}_k\label{RxSignalbeforeADC_unsynced}\ 
\end{IEEEeqnarray}
where $\epsilon$, $\phi$, and $\Omega$ are the channel parameters denoting the timing, phase, and frequency offset, respectively. Moreover, we additionally introduced a random phase dither $\varphi_k$ which will be discussed below. Note that for the signal model in (\ref{RxSignalbeforeADC_unsynced}) we have, differently to the sections before, assumed that the receive filter $h_{\textrm{Rx}}(t)$ has a rectangular transfer function with one-sided bandwidth $W_r=\frac{1}{2T_s}$. This enables an analytic treatment, as we adapted the receive filter bandwidth to the sampling rate such that the circularly symmetric complex Gaussian noise samples $\mathsf{n}_k$ are mutually independent. In general the samples of the noise process $\{\mathsf{n}_k\}$ will be correlated, as the receive filter bandwidth should be chosen to match the bandwidth of the transmit signal and we consider temporal oversampling. Nevertheless, the study of the white noise case gives important insights for the practical case of a colored noise process as shown in \cite{Schluter_LS_phase}. In addition, we assume for the channel parameter estimation that the transmit symbols $\mathsf{x}_n$ are i.i.d.\ and zero-mean. Thus, we can defined the symbol energy as $E_s=\mathrm{E}[|\mathsf{x}_n|^2]\int_{-\infty}^{\infty}|h(t)|^2\mathrm{d}t$.

A first step to the design of channel parameter estimation algorithms is to analyze the impact of the 1-bit quantization on the Fisher information (FI) and the Cram\'er-Rao lower bound (CRLB), which give insights on the performance limits of actual channel parameter estimation algorithms. 

\begin{figure}
\center	\includegraphics[width=0.48\textwidth]{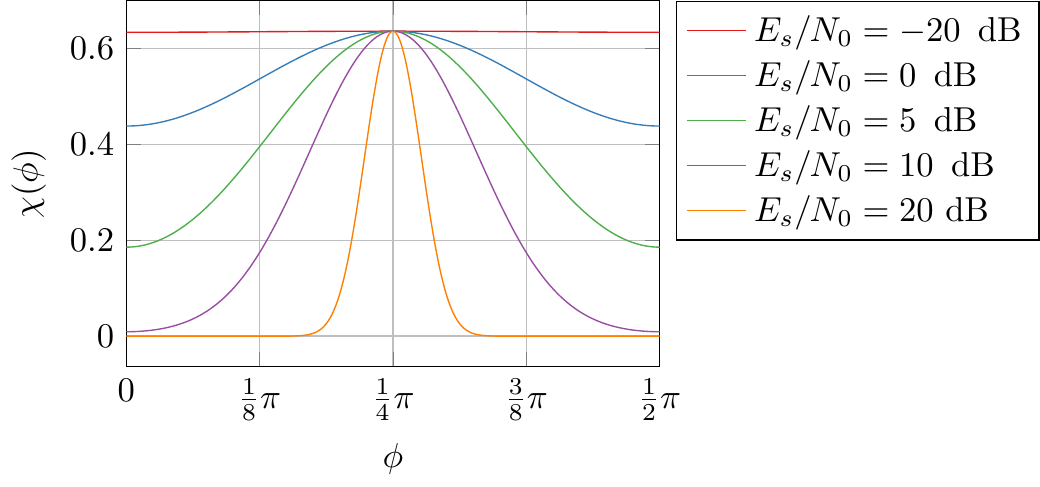}\vspace{-0.2cm}
	\caption{Ratio $\chi(\phi)$ between the FI of the 1-bit quantized received signal and the FI of the unquantized received signal for a single QPSK symbol $\mathsf{x}_k$ (i.e., $N=1$), $M=1$, and $\epsilon$ and $\Omega$ known to be zero, $\varphi_k=0$.\cite{schluterbounds}}
	\label{fig:FI-Phase_over_phi}
\end{figure}

We derived the FI matrix $\vec{F}_{\vec{\mathsf{y}}}$ for the phase $\phi$ and the frequency $\Omega$, considering a known timing error $\epsilon$ and known data symbols $\{\mathsf{x}_n\}$ \cite{schluterbounds}. Amongst other things, it was found that the parameter estimation accuracy is highly dependent on the phase of the receive signal, unlike in the unquantized case where the FI is phase independent. If the samples $\mathsf{r}_k$ are close to the decision boundary in the complex plane, the FI is high, but when the samples lie within the middle of a quadrant, the FI is low. The higher the SNR, the more pronounced is this effect. This is intuitive, since in the noiseless case for every sample $\mathsf{r}_k$ the same 1-bit quantized measurement $\mathsf{y}_k$ would be observed, which results in a poor estimation performance. In Fig.~\ref{fig:FI-Phase_over_phi} this effect is illustrated by the loss function
\begin{equation}
	\chi(\phi) = \frac{\left[\vec{F}_{\bm{\mathsf{r}}}\right]^{-1}_{\phi \phi}}{\left[\vec{F}_{\bm{\mathsf{y}}}\right]^{-1}_{\phi \phi}}
\label{eq:FI_ratio}
\end{equation}
for the phase estimation of a single QPSK symbol under the assumption that $\epsilon$ and $\Omega$ are known to be zero. We see that the phase dependency is high for high SNR and vanishes for low SNR, where the loss factor approaches $\frac{2}{\pi}$, a fact well known \cite{Mezghani2010,stein2015asymptotic}.

\begin{figure}
\center	\includegraphics[width=0.48\textwidth]{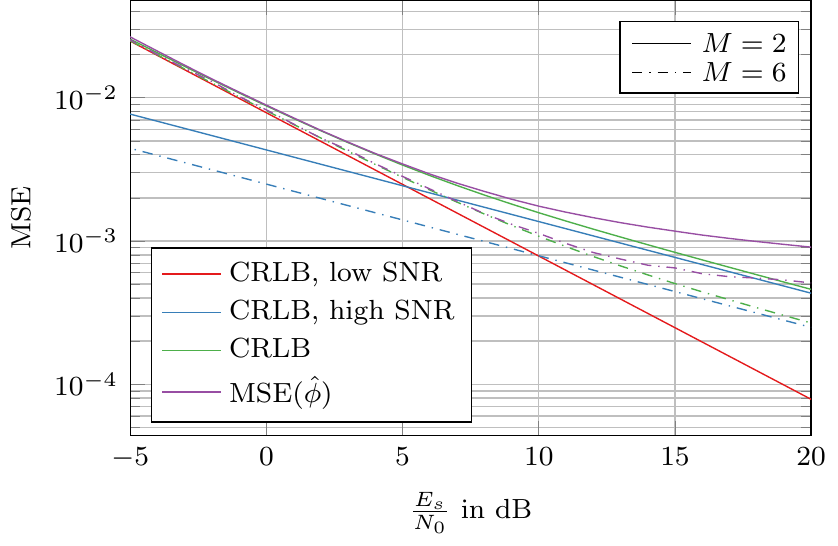}\vspace{-0.1cm}
	\caption{High and low SNR bounds on the CRLB $[\mathbf{F}_{\bm{\mathsf{y}}}]_{\phi\phi}^{-1}$ and the MSE of the LS phase estimator; $h(t)$ is a root raised cosine pulse with rolloff factor 0.2, $M_{\textrm{Tx}}=1$, i.e., there is no FTN signaling, $N=100$, \cite{schluterbounds,Schluter_LS_phase}.}
	\label{fig:CRLB_Phase_Bounds}
\end{figure}

This observation motivates to add a uniform phase dither $\varphi_k\in\mathcal{U}[0,2\pi]$, see (\ref{RxSignalbeforeADC_unsynced}). A similar behavior can be achieved in a system implementation by sampling the received signal at a low intermediate frequency (IF) that is irrational when normalized on the sampling rate. For this case we derived tight closed form upper bounds on the FI for the high and the low SNR case \cite{schluterbounds}. These bounds show that compared to the unquantized case many properties are preserved under 1-bit quantization combined with phase dithering. The major difference is that for high SNR the FI for phase and frequency only grows in the order of the square root of $\frac{E_s}{N_0}$, but in return increases with the square root of the oversampling factor $M$. Differently, in the unquantized case the FI grows linearly with $\frac{E_s}{N_0}$ and oversampling has no effect. As the Cram\'er-Rao lower bound (CRLB), which is a lower bound on the estimation error variance, is given by the inverse of the FI, the upper bounds on the FI give lower bounds on the CRLB. Moreover, we observed that as in the unquantized case the cross term $[\mathbf{F}_{\bm{\mathsf{y}}}]_{\phi\Omega}$ can be neglected for sufficiently large $N$ such that the phase estimation error variance is lower-bounded by  $[\mathbf{F}_{\bm{\mathsf{y}}}]_{\phi\phi}^{-1}$. Exemplarily, we get the following lower bounds on the CRLB for the phase estimation variance which are shown in Fig.~\ref{fig:CRLB_Phase_Bounds}
\begin{IEEEeqnarray}{rCL}
\textrm{CRLB}(\phi)&\ge&\!\left\{\!\!\!\begin{array}{ll}
\left(\frac{4}{\pi}\frac{E_s}{N_0}\right)^{-1}\frac{1}{N} & \textrm{for low SNR}\\
\left(\frac{2c_1}{\sqrt{2\pi^3c_2}}\sqrt{\frac{E_s}{N_0}}\right)^{-1}\frac{1}{N\sqrt{M}} & \textrm{for high SNR}
\end{array}\right.  
\end{IEEEeqnarray}
Here $c_1$ and $c_2$ are some constants given in \cite{schluterbounds}. For comparison Fig.~\ref{fig:CRLB_Phase_Bounds} also shows the actual CRLB. 

In \cite{Schluter_LS_phase} we presented a phase estimator based on the least squares estimator (LSE), which assumes that the timing $\epsilon$ and the data $\{\mathsf{x}_n\}$ are known. We assume that there is a coarse frequency synchronization in the analog stages such that the unknown phase rotation of the receive signal can assumed to be constant over the observation interval of the estimator, i.e., $\Omega \approx 0$. Instead of directly applying the least squares criterion to the phase rotation, we apply it to the phasor $\exp(j\phi)$ such that the signal model without 1-bit quantization is linear. Taking the argument of the estimated phasor gives
\begin{IEEEeqnarray}{rCl}
    \hat{\mathsf{\phi}} = \mathrm{arg}\left( \sum_k \mathsf{u}_k^* e^{-j\varphi_k} \mathsf{y}_k \right).
\end{IEEEeqnarray}
Although this is a very simple linear estimator, we could prove that it is consistent due to uniform phase dithering. Thus, for large $N$ we can compare its mean squared error (MSE) against the CRLB as shown in Fig.~\ref{fig:CRLB_Phase_Bounds}. 

We observe that at low SNR the MSE of the LSE achieves the CRLB. However, at high SNR the MSE is converging to a fixed value. This behavior has a simple explanation. As pointed out before, it is known that at low SNR, the 1-bit quantized system behaves very similar to the unquantized system, except of a constant performance loss of $\frac{2}{\pi}$ \cite{Mezghani2010}. Moreover, it is known that the LSE is in fact the maximum likelihood estimator (MLE) for AWGN channels, where the MLE asymptotically achieves the CRLB. Thus, at low SNR the LSE asymptotically achieves the CRLB for the 1-bit quantized channel. However, at high SNR the nonlinear effect of the 1-bit quantizer becomes much more dominant and, thus, the LSE does not achieve the CRLB, which is likely due to its simple linear nature. Furthermore, we observe that as predicted by the CRLB, oversampling improves the performance of the LSE. For higher oversampling factors its MSE departs from the CRLB at a higher $\frac{E_s}{N_0}$ and converges to a lower value.

One key observation regarding the channel parameter estimation is the fact that it performs best in case the observation samples lie close to a decision boundary, see Fig.~\ref{fig:FI-Phase_over_phi}. This can be achieved by an optimal phase dithering in the analog domain based on feedback from the digital side. On the other hand, in order to maximize the achievable rate the phase rotation must be eliminated such that the QPSK symbols are rotated back to their original positions and, thus, the noise sensitivity is minimized. Also this can be achieved by a feedback from the digital into the analog domain. Note that the optimal phase rotation for parameter estimation and data detection are in contradiction to each other. 

In terms of the phase estimation we have argued that uniform phase dithering, e.g., generated by an irrational normalized low IF, is a suitable approach to enable phase estimation. However, it only gives the average performance. Likewise, for timing synchronization a similar behavior can be achieved by choosing an irrational ratio between sampling and signaling rate \cite{Schluter2018_timing}. These approaches have the advantage that they do not need any feedback from the digital domain back to the analog frontend of the receiver and, thus, are simple to implement. They can also be used in the communication phase, but again will only give the average performance. In order to achieve optimal performance, the phase and timing offset must be compensated, e.g., the QPSK symbols must be rotated back to their original position.

We have shown in \cite{Schluter2018_timing} that oversampling can compensate for the communication performance loss of a timing shift. Also in case of channel parameter estimation, we have shown that the performance gap between uniform dithering and optimal dithering can be closed by oversampling \cite{schluterbounds}. However, the larger the SNR, the faster the receiver has to sample to account for the performance loss, see Fig.~\ref{fig:CRLB_Phase_Bounds}. 

The tradeoffs between, on the one hand, the performance of channel parameter estimation and communication, and, on the other hand, the energy efficiency and implementation complexity with different system architectures using oversampling, low IF sampling based dithering, and feedback based dithering is an open question.

\section{Continuous Phase Modulation}
By relieving linearity requirements at transmitter and receiver jointly, CPM and 1-bit quantization are a promising combination for energy efficient communication at high carrier frequencies and large bandwidths. When combining this approach with oversampling and sampling at an IF, sequences of zero-crossings (more precisely, axis crossings in the IQ-plane) can be designed that allow to increase the spectral efficiency beyond \SI{2}{bit/s/Hz}. Achievable rate and spectral efficiencies of this approach were analyzed in \cite{Landau18TWC} for a limited set of IFs, and in \cite{Bender2019} the impact of the IF was investigated and bit error rate (BER) results were given. This section provides an overview of the results.

\subsection{CPM System Model}
A complex baseband CPM-signal is described as
\begin{IEEEeqnarray}{rCL}
	\mathsf{c}(t) = \sqrt{\frac{2 E_s}{T}} e^{j \upphi(t)}. \label{eq:tx_signal}
\end{IEEEeqnarray}
The phase term $\upphi(t)$ is given by 
\begin{IEEEeqnarray}{rCL}
	\upphi(t) = 2 \pi h_\mathrm{cpm} \sum\nolimits_{n=0}^{N-1} \upalpha_n q(t-nT,T) +\phi_0 \label{eq:phase}
\end{IEEEeqnarray}
where the $\upalpha_n \in \mathcal{A} = \{\pm 1, \pm 3, ... , \pm(M_\mathrm{cpm}-1)\}$ are the CPM-symbols with alphabet size $M_\mathrm{cpm}$, $\phi_0$ is the initial phase at $t=0$,  and $h_\mathrm{cpm}$ is the modulation index. The function $q(t,T)$ is the phase smoothing response. We consider continuous phase frequency shift keying (CPFSK), which has a linear characteristic
\begin{IEEEeqnarray}{rCL}
	q(t,T) &= \begin{cases}
		0, & \text{if } t \le 0\\
		\frac{t}{2 L_\mathrm{cpm} T}, & \text{if } 0 < t \le L_\mathrm{cpm} T\\
		\frac12, & \text{if } t > L_\mathrm{cpm} T
	\end{cases}.\label{eq:phase-smoothing-response}
\end{IEEEeqnarray}
Here, $L_\mathrm{cpm}$ is the memory of the CPM-symbol in terms of the symbol duration $T$. We assume $h_\mathrm{cpm}=\frac{1}{M_\mathrm{cpm}}$, $L_\mathrm{cpm}=1$, $\phi_0 = \frac{\pi}{M_\mathrm{cpm}}$, and $M_{\text{cpm}}$ to be even. Furthermore, $M_\mathrm{Tx}=1$ and i.u.d.\ input symbols $\upalpha_n$ are assumed. 

The continuous-time channel input signal is generated by 
\begin{IEEEeqnarray}{rCL}
	\mathsf{x}(t) = \Re \{(h(t) \ast \mathsf{c}(t)) e^{j 2 \pi f_0 t}\}. \label{eq:cpm_tx_signal}
\end{IEEEeqnarray}
The complex received signal, down-converted to an IF, is
\begin{IEEEeqnarray}{rCL}
	\mathsf{z}(t) = g(t) \ast (\mathsf{c}(t) e^{j 2 \pi f_\mathrm{IF} t})+h_\mathrm{Rx}(t) \ast \mathsf{n}(t). \label{eq:cpm_rx_signal}
\end{IEEEeqnarray}
The filters $g(t)$ and $h_\mathrm{Rx}(t)$ can be bandpass-filters if $f_\mathrm{IF}>0$. The oversampled 1-bit ADC output signal is then generated according to (\ref{eq:Qunatizer1bit}) with $\mathsf{r}_k =\mathsf{z}(k T_s)$.

\subsection{Achievable Rate \& Noise and Distortion Free Analysis}
\begin{figure}
	\includegraphics[width=0.48\textwidth]{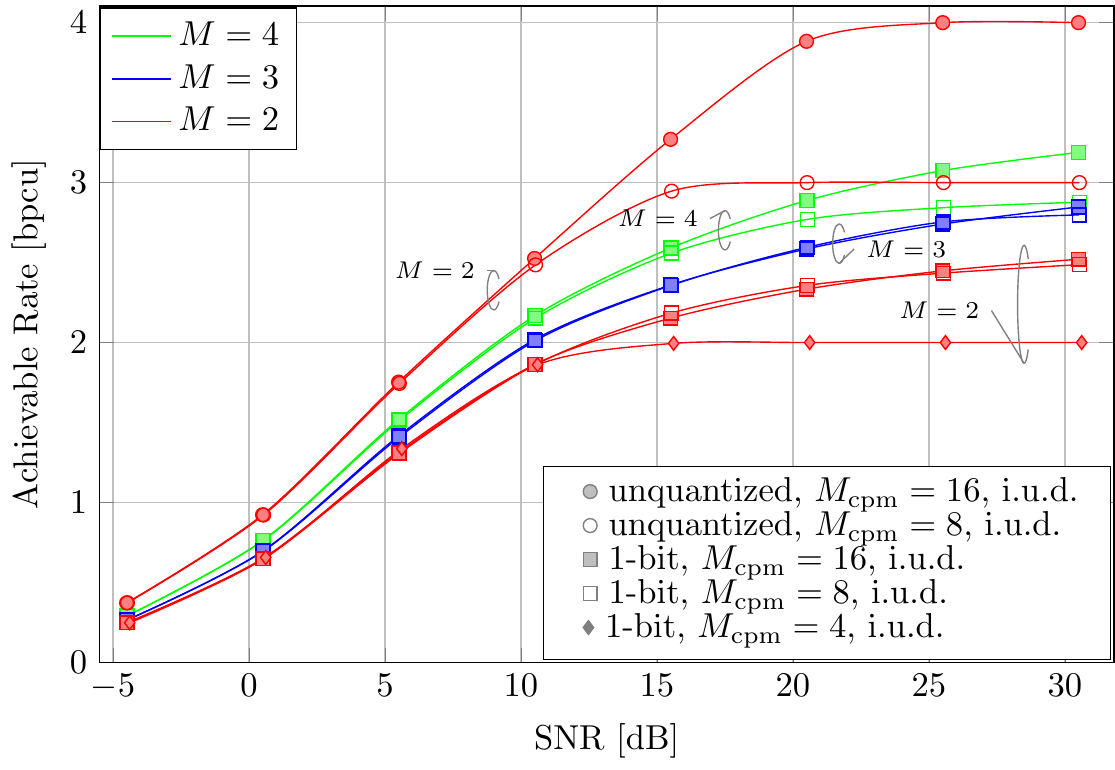}\vspace{-0.1cm}
	\caption{Achievable rate depending on $M$ and $M_\mathrm{cpm}$ for $f_\mathrm{IF}=\Delta f$ and $n_\mathrm{IF}=0$, $h_\mathrm{Rx}(t) = \sqrt{\frac{2}{T} \rect \left(\frac{2 t}{T}\right) e^{j 2 \pi (\Delta f + \frac{n_\textrm{IF}}{T})t}}$, $h(t)=\delta(t)$ \cite{Landau18TWC}.}
	\label{fig:AchRate_Lukas}
\end{figure}
\begin{figure}
	\includegraphics[width=0.48\textwidth]{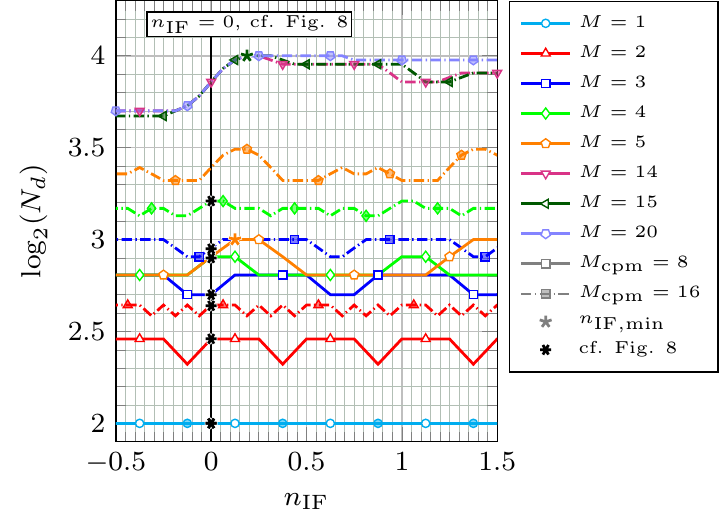}\vspace{-0.1cm}
	\caption{Number of distinguishable bits $\log_2(N_d)$ in a noise and distortion free scenario for CPFSK depending on $M$ and $f_\mathrm{IF}$.}
	\label{fig:CPM_number_dist_paths}
\end{figure}
The phase $\upphi(t)$ in (\ref{eq:phase}) can be represented in a trellis diagram. Using the definition of $\mathcal{A}$ given above, this trellis is not time invariant. It can be transformed into a time invariant trellis \cite{Rimoldi88} by adding a frequency offset, i.e.,  $\uppsi(t)=\upphi(t)+2\pi\Delta f t$ with $\Delta f= \frac{h_\mathrm{cpm}(M_\mathrm{cpm}-1)}{2 T}$, which corresponds to an IF. For analyzing the impact of the IF, we adopt the notation
\begin{eqnarray}
{f}_{\mathrm{IF}}=\Delta f + \frac{n_\mathrm{IF}}{T}.
\end{eqnarray}
Fig.~\ref{fig:AchRate_Lukas} shows the achievable rates in bits per channel use (bpcu) for $n_\mathrm{IF}=0$, where the results are obtained using the techniques presented in Section~\ref{subsec:AchRate}. One can observe the benefit of oversampling w.r.t.\ the achievable rate. In Fig.~\ref{fig:CPM_number_dist_paths}, we choose $n_\mathrm{IF}=c h_\mathrm{cpm},~ c \in \mathbb{Z}$. This retains the property of a time invariant trellis, since the additional phase turn is always an integer multiple of the angle between two constellation points. We depict the number of bits $\log_2(N_d)$ that can be resolved per $T$ for a given modulation order, oversampling factor, and IF in an ideal transmission scenario (no noise, no filter distortion). Here, $N_d$ is the average number of paths per $T$ in the trellis that remain distinguishable under 1-bit quantization. It can be seen that the achievable rate per channel use in the high-SNR limit from Fig.~\ref{fig:AchRate_Lukas} matches the theoretical limit approximately; variations are due to filtering. When increasing the IF beyond $\Delta f$, all symbols can be resolved in this ideal scenario given oversampling factors $M=5$ and $M=14$\footnote{Note, that in the presence of noise odd oversampling factors are to be preferred as this avoids samples on the decision threshold of the quantizer.}, for 8- and 16-CPM, respectively. The minimum IF for resolving all symbols with a sufficiently high $M$ is
\begin{eqnarray}
    n_\textrm{IF,min}=\left\lceil\frac{1}{4 h_\mathrm{cpm}}-1\right\rceil h_\mathrm{cpm}.
\end{eqnarray}
Furthermore, $\log_2(N_d)$ is periodic over $n_\textrm{IF}$ and can decrease again for $n_\textrm{IF} > n_\textrm{IF,min}$ if different trellis paths become indistinguishable again. 

BER results for $f_{\mathrm{IF}} = \Delta f + \frac{1}{4T}$ using root raised cosine filters are given in \cite{Bender2019}, where also the possibility of adapting the phase smoothing response w.r.t.\ 1-bit quantization is studied.\looseness-1

\section{Conclusion}
In the design of energy efficient wireless communication systems with data rates beyond \SI{100}{\giga bit/s} the ADC poses a bottleneck in terms of the power consumption. Using energy efficient 1-bit quantization with temporal oversampling is considered to be a suitable solution to this problem. Using receivers with 1-bit quantization and temporal oversampling the information must be encoded in the temporal distance of the zero crossings of the transmit signal, i.e., we have to modulate the zero crossings. This means that information is conveyed in the time domain instead of the amplitude domain of the transmit signal. In the present paper, we have shown that runlength coding in combination with FTN signaling is a suitable signaling approach. We have discussed different aspects of this signaling scheme and the corresponding communication system design. We studied its spectral efficiency, showing that it can be significantly increased depending on the FTN signaling rate. Regarding the actual system design we discussed mapping, demapping, and equalization of RLL sequences and studied aspects of the channel parameter estimation. We have observed that additional oversampling w.r.t.\ the signaling rate improves the coded bit error rate as well as the channel parameter estimation. Moreover, phase dithering (implemented by an appropriate low IF sampling) significantly improves the asymptotic behavior of the channel parameter estimators. Finally, the use of continuous phase modulation in case of temporally oversampled 1-bit quantization has been discussed, having the additional advantage of enabling the use of efficient power amplifiers at the transmitter. This study shows possible gains in the achievable rate when sampling at a carefully chosen low IF with a sufficiently high oversampling factor.

\bibliographystyle{IEEEtran}
\bibliography{bib_1bit}

\end{document}

%% file: fig_eval_bler.tex
\definecolor{mycolor1}{rgb}{0.00000, 0.4470, 0.7410}%
\definecolor{mycolor2}{rgb}{0.85000, 0.32500, 0.09800}%
\definecolor{mycolor3}{rgb}{0.9290,0.6940, 0.1250}%
\definecolor{mycolor4}{rgb}{0.4940, 0.1840, 0.5560}%
\definecolor{mycolor5}{rgb}{0.4660, 0.6740, 0.1880}%
\definecolor{mycolor6}{rgb}{0.3010, 0.7450, 0.9330}%
\definecolor{mycolor7}{rgb}{0.6350, 0.0780, 0.1840}%
\begin{tikzpicture}[font=\footnotesize]
	\begin{axis}[%
	width=1.0\columnwidth,
	height=0.6\columnwidth,
	xmin=7,
	xmax=13,
	xlabel={$E_\mathrm{b}/N_0$ [\SI{}{\decibel}]},
	ymode=log,
	ymin=1e-03,
	ymax=1,
	yminorticks=true,
	ylabel={BLER},
	axis background/.style={fill=white},
	axis x line*=bottom,
	axis y line*=left,
	xmajorgrids,
	xminorgrids,
	ymajorgrids,
	yminorgrids,
	legend pos=south west,
	]
		\addlegendimage{black, line width=0.75pt}
		\addlegendentry{$M=1$}
		\addlegendimage{black, dashed, line width=0.75pt}
		\addlegendentry{$M=2$}
		\addlegendimage{black, dotted, line width=0.75pt}
		\addlegendentry{$M=3$}
		\addlegendimage{mycolor1, mark=x, line width=0.75pt}
		\addlegendentry{$R_\mathrm{FEC}=8/9$}
		\addlegendimage{mycolor3, mark=square, line width=0.75pt}
		\addlegendentry{$R_\mathrm{FEC}=4/5$}
		\addlegendimage{mycolor2, mark=o, line width=0.75pt}
		\addlegendentry{$R_\mathrm{FEC}=3/4$}
		
		\addplot [color=mycolor1, mark=x, mark options={solid, mycolor1}, line width=0.75pt]
		table[x index=0,y index=1]{data/data_asilomar_bler_block_r8-9_mtx2_mrx2-4-6_d1.dat};
		\addplot [color=mycolor1, dashed, mark=x, mark options={solid, mycolor1}, line width=0.75pt]
		table[x index=0,y index=2]{data/data_asilomar_bler_block_r8-9_mtx2_mrx2-4-6_d1.dat};
		\addplot [color=mycolor1, dotted, mark=x, mark options={solid, mycolor1}, line width=0.75pt]
		table[x index=0,y index=3]{data/data_asilomar_bler_block_r8-9_mtx2_mrx2-4-6_d1.dat};
		
		\addplot [color=mycolor3, mark=square, mark options={solid, mycolor3}, line width=0.75pt]
		table[x index=0,y index=1]{data/data_asilomar_bler_block_r4-5_mtx2_mrx2-4-6_d1.dat};
		\addplot [color=mycolor3, dashed, mark=square, mark options={solid, mycolor3}, line width=0.75pt]
		table[x index=0,y index=2]{data/data_asilomar_bler_block_r4-5_mtx2_mrx2-4-6_d1.dat};
		\addplot [color=mycolor3, dotted, mark=square, mark options={solid, mycolor3}, line width=0.75pt]
		table[x index=0,y index=3]{data/data_asilomar_bler_block_r4-5_mtx2_mrx2-4-6_d1.dat};
		
		\addplot [color=mycolor2, mark=o, mark options={solid, mycolor2}, line width=0.75pt]
		table[x index=0,y index=1]{data/data_asilomar_bler_block_r3-4_mtx2_mrx2-4-6_d1.dat};
		\addplot [color=mycolor2, dashed, mark=o, mark options={solid, mycolor2}, line width=0.75pt]
		table[x index=0,y index=2]{data/data_asilomar_bler_block_r3-4_mtx2_mrx2-4-6_d1.dat};
		\addplot [color=mycolor2, dotted, mark=o, mark options={solid, mycolor2}, line width=0.75pt]
		table[x index=0,y index=3]{data/data_asilomar_bler_block_r3-4_mtx2_mrx2-4-6_d1.dat};
		
	\end{axis}
\end{tikzpicture}%